\documentclass[letter,twocolumn]{jpsj3} 
\usepackage{physics} 
\usepackage{txfonts}
\usepackage{xcolor}
\usepackage{amsmath,amssymb,mathtools} 
\usepackage[capitalize]{cleveref} 
\crefrangeformat{equation}{(#3#1#4)-(#5#2#6)}

\title{
  Andreev-like Reflection in the Pfaffian Fractional Quantum Hall Effect
} 

\author{ 
  Ryoi Ohashi$^1$,
  Ryota Nakai$^2$,
  Takehito Yokoyama$^3$, 
  Yukio Tanaka$^1$,
  and Kentaro Nomura$^2$ 
} 
\inst{ 
  $^1$Department of Applied Physics, Nagoya University, Nagoya 464-8603, Japan\\ 
  $^2$Department of Physics, Kyushu University, Fukuoka 819-0395, Japan\\ 
  $^3$Department of Physics, Tokyo Institute of Technology, Meguro, Tokyo 152-8551, Japan 
} 

\abst{ 
  We studied the tunnel transport between the edge of a Pfaffian fractional quantum Hall state and that of an integer quantum Hall state. 
  Based on the duality argument between the strong and weak tunnelings, we found that an Andreev-like reflection appeared in the strong tunneling regime. 
  We calculated the charge conductance in the weak and strong tunneling regimes for the low-voltage limit. 
  In the weak tunneling limit, $\dd{I}/\dd{V}$ was proportional to $V^{1/\nu}$ with bias voltage $V$ and $\nu=1/2$. 
  By contrast, in the strong tunneling limit, $\dd{I}/\dd{V}$ was expressed by $(e^{2}/h)2\nu/(1+\nu)$ with a correction term. 
  We expect that this condition can be realized experimentally at the point contact between a fractional quantum Hall state with $\nu=5/2$ and an integer quantum Hall state with $\nu=3$.
}

\begin{document} 
\maketitle 

The fractional quantum Hall (FQH) effect is a phenomenon in which Hall conductivity takes fractionally quantized values.\cite{klitzing_1980, tsui_1982,willett_1987a} 
The fractional values are characterized by the filling factor $\nu$, which is the ratio of the number of fluxes to the number of electrons. 
Among these, the most fundamental is the FQH effect, whose filling factor is of the form $\nu=1/(2k+1)$ with an odd-denominator. 
For this state, a trial wave function was introduced by Laughlin,\cite{laughlin_1983} which was constructed at a single Landau level. 
The above states are limited to odd-denominator filling factors by the restriction of preserving the antisymmetry of the fermions. 
However, Moore and Read proposed a wave function for FQH states with even denominators and succeeded in describing a wider variety of FQH states.\cite{moore_1991} 
This Pfaffian FQH state is associated with the correlation function of the two-dimensional Ising conformal field theory. 
It is suggested that this state exists at the $N=1$ Landau level with $\nu=5/2$.\cite{greiter_1991,greiter_1992} 

At the edge of the FQH states, there is a gapless state owing to the Chern--Simons field defined in the bulk effective field theory.\cite{wen_2004} 
This gapless state can be described as a Luttinger fluid by the bosonization technique, and quantum transport phenomena have been analyzed using edge states. 
Charge transport can be analyzed by using point contact.\cite{kane_1992,imura_1998,nomura_2001,ito_2012,milliken_1996,radu_2008,dolev_2008,chang_2003}
There are two types of processes: quasiparticle tunneling through the bulk of a fractional quantum Hall liquid (FQHL) and electron tunneling between the two counter-propagating edges of the two FQHLs. 
These two processes are related through the duality both in the Laughlin and Pfaffian states.\cite{kane_1992,nomura_2001,fendley_2007a} 
In addition, a point contact between the Laughlin FQH state and an integer quantum Hall (IQH) system was theoretically proposed.\cite{sandler_1998} 
This system also exhibits Andreev-like reflections in the strong coupling limit. 
Andreev-like reflections in this configuration are analogous to Andreev reflections in superconducting junctions,\cite{andreev_1965,blonder_1982} where $k+1$ quasiparticles of fractional charge $e^{\ast}=\nu e=e/(2k+1)$ are injected on the FQH side and $k$ quasiholes with fractional charge are returned as reflections; thus, the process is $(k+1)e^{\ast}-k(-e^{\ast})=(2k+1)e^{\ast}=e$, which is the amount of transmission to the IQH side. 
The theoretical proposal of this Andreev-like reflection process was recently examined experimentally.\cite{hashisaka_2021} 

Recent studies have focused on anyon interferences and collisions in the Laughlin FQH state.\cite{rosenow_2016,bartolomei_2020,nakamura_2020}, while the $\nu=5/2$ FQH system has also been observed in the measurements of transport phenomena at point contacts and fractional charge by shot noise.\cite{radu_2008,dolev_2008}
To extend the experiments performed in the Laughlin FQH state to the $\nu=5/2$ FQH state, it is essential to clarify the fundamental transport phenomena.
Moreover, quasiparticle excitations in the Pfaffian FQH state exhibit non-Abelian statistics, and thus attracted considerable attention from the viewpoint of quantum computation.\cite{nayak_2008,kitaev_2003,mong_2014a} 
Extracting this non-Abelian statistics via transport phenomena can potentially facilitate the development of physics.

In this letter, we extend the Andreev-like reflections to the Pfaffian FQH state and discuss their transport phenomena. 
As a result, we confirmed the existence of two reflection processes in the strongly-coupled limit: a normal reflection between quasiparticles and electrons with $2e^{\ast}=\nu^{(1)}e$ and an Andreev-like reflection by quasiparticles with $e^{\ast}=\nu^{(1)}e/2$. 
Here, $\nu^{(1)}=1/2$ is the filling factor at the $N=1$ Landau level and in the following $\nu^{(1)}$ is simply denoted by $\nu$. 

\begin{figure}[htb] 
  \centering 
  \includegraphics[width=0.8\linewidth]{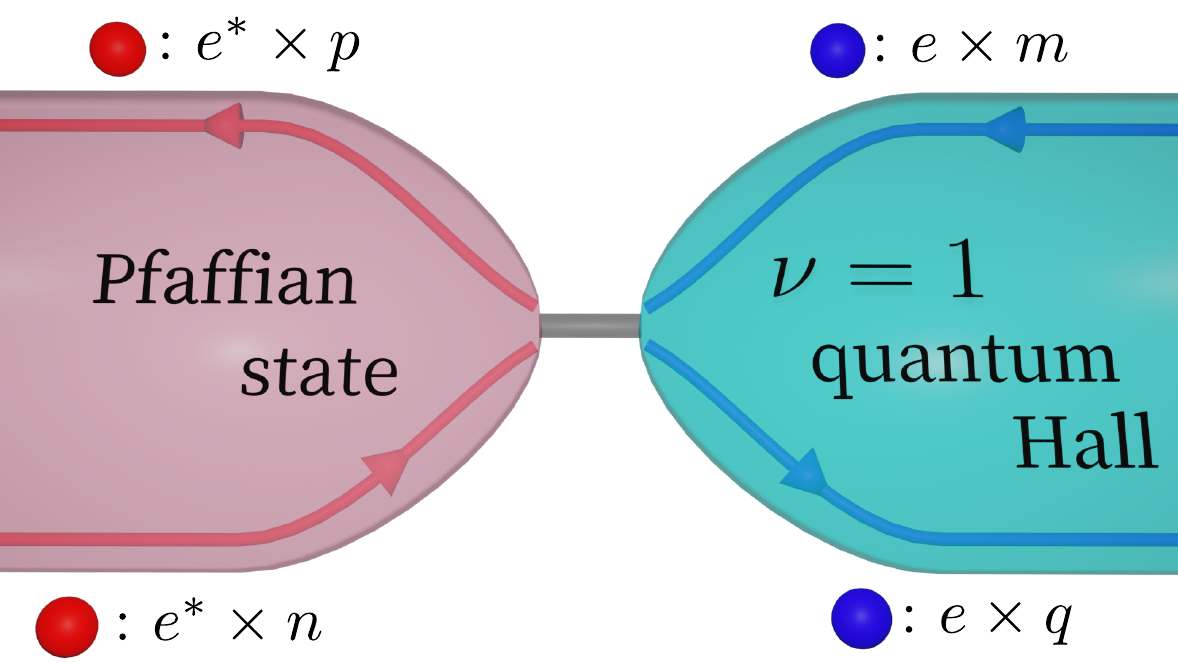} 
  \caption{(Color online)
  Schematic image of the 
  scattering process: $m$ incoming electrons and $n$ incoming quasiparticles are scattered into $q$ and $p$ outgoing electrons and quasiparticles, respectively, through electron tunneling. 
  Quasiparticles~\cref{eqn:qp1,eqn:qp2,eqn:qp3,eqn:qp4} are applied according to the number of $n$ and $p$. 
  The elementary charge is $e^{\ast} = e/4$.
  } 
  \label{fg: geometry} 
\end{figure} 
We consider a $\nu = 1/2$ Pfaffian FQHL $/$ integer quantum Hall liquid (IQHL) junction with a point contact at $x = 0$, as shown in~\cref{fg: geometry}. 
The low-energy excitations of the FQH effect side are the modes that localized near the edges because the bulk quantum Hall state has an energy gap. 
The edge states of the Pfaffian state are described by conformal field theory with a central charge $c = 1 + \frac{1}{2}$.\cite{moore_1991} 
The $c = 1/2$ part of the edge modes belongs to the same universality class as the critical point of the two-dimensional Ising model. 
The primary fields are the real fermion $\chi$, spin operator $\sigma$, and unit operator $ 1 $.\cite{belavin_1984,francesco_1996_conformal} 
These modes are related to the pair breaking and $h/2e$ quantum vortex excitations in the bulk.\cite{moore_1991,milovanovic_1996} 
The $c = 1$ part describes the Laughlin-type excitations, which correspond to the free boson theory.\cite{wen_1990,wen_1992,wen_2004} 
We denote this bosonic field by $\phi_{c}$. 
On the other hand, we treat the IQHL as a one-dimensional chiral Fermi liquid that appears on the edge state of a $\nu = 1$ IQHL, for which the bosonized theory is also applicable. 
By writing the latter bosonic field as $\phi_f$, the edge theory can be written as \cite{moore_1991} 
\begin{align} 
  H = \int\dd{x}\qty[-\frac{i\varv}{2}\chi\partial_{x}\chi + \frac{\varv}{4\pi}\qty(\partial_{x}\phi_{c})^2 + \frac{\varv}{4\pi}\qty(\partial_{x}\phi_{f})^2] ,
\end{align} 
where the fields $\phi_{c}$ and $\phi_{f}$ satisfy the commutation relations 
\begin{align} 
  \comm{\phi_{I}(x)}{\phi_{J}(x^{\prime})} 
  = i\pi\delta_{IJ}\mathrm{sgn}(x-x^{\prime}) 
\end{align} 
with $I,J=c,f$. 
For simplicity, we assume that the velocities of the edge modes are equal to $\varv$. 
On the edge of the Pfaffian state side, the electron annihilation operator is given by\cite{moore_1991}
\begin{align} 
  \psi_{\mathrm{e}c}=\chi e^{i\phi_{c}/\sqrt{\nu}}. 
\end{align} 
Here, the lowest Landau level contribution is neglected. In the Pfaffian FQH state, there are four types of charged quasiparticles, characterized by the following operators\cite{moore_1991,fendley_2007a, fradkin_2013} 
\begin{align} 
  \psi_{\nu e,1}          &= 1      e^{i \sqrt{\nu}\phi_{c}  },\label{eqn:qp1}\\
  \psi_{\nu e,\chi}       &= \chi   e^{i \sqrt{\nu}\phi_{c}  },\label{eqn:qp2}\\
  \psi_{\nu e/2,\sigma}   &= \sigma e^{i \sqrt{\nu}\phi_{c}/2},\label{eqn:qp3}\\
  \psi_{3\nu e/2,\sigma}  &= \sigma e^{i3\sqrt{\nu}\phi_{c}/2}.\label{eqn:qp4}
\end{align} 
Each exponential is assumed to be normal ordered. 
\Cref{eqn:qp1} corresponds to the Laughlin-type quasiparticles while the rest of the operators are specific to the Pfaffian state. 
As indicated by the indices of the above operators, the charges of these quasiparticles are $\nu e$, $\nu e$, $\nu e/2$, and $3\nu e/2$, respectively. 
On the IQHL side, the electron operator is expressed as $\psi_{\mathrm{e}f} = e^{i\phi_{f}}$. 
Then, the electron tunneling at $x = 0$ is given by the Hamiltonian 
\begin{align} 
  \begin{split} 
    H_\mathrm{T} &= \int\dd{x} t\delta(x)\qty(\psi_{\mathrm{e}c}^{\dagger}\psi_{\mathrm{e}f}+\mathrm{h.c.}) \\ 
    &= \int\dd{x} \Gamma\delta(x)\chi\cos(\frac{1}{\sqrt{\nu}}\phi_{c}-\phi_{f}). 
  \end{split} 
\end{align} 

This model has two fixed points: one is a stable fixed point at $\Gamma = 0$ (weak coupling limit) and the other is an unstable fixed point at $\Gamma = \infty$ (strong coupling limit). 
The first case is trivial, because it corresponds to two decoupled systems. 
For analyzing the second fixed point, the duality symmetry is useful, and for the edge theory of the $\nu = 1/(2k + 1)$ Laughlin states this symmetry is well known.\cite{kane_1992,saminadayar_1997} 

We consider the scattering process shown in~\cref{fg: geometry}. 
In the incoming state, we have $m$ electrons on the Fermi liquid side with total charge $me$ and $n$ quasiparticles on the Pfaffian FQHL side with total charge $ne^{\ast}$.
The outgoing state has $q$ electrons with charge $qe$ and $p$ quasiparticles with charge $pe^{\ast}$. 
The probability amplitude for such process is proportional to \cite{sandler_1998} 
\begin{align} 
  \expval{ 
  \eta_{p}^\mathrm{out}e^{i\frac{\sqrt{\nu}}{2}p\phi_{c}^\mathrm{out}} 
  e^{iq\phi_{f}^\mathrm{out}} 
  \eta_{n}^\mathrm{in}e^{-i\frac{\sqrt{\nu}}{2}n\phi_{c}^\mathrm{in}} 
  e^{-im\phi_{f}^\mathrm{in}}, 
  } 
  \label{eqn:expval_1} 
\end{align} 
where the $\eta$ operators are $\sigma$, $\chi$, or $1$ in the $c = \frac{1}{2}$ part, which is determined by $p$ and $n$ based on~\cref{eqn:qp1,eqn:qp2,eqn:qp3,eqn:qp4}.

\subsection*{Weak Coupling Limit} 
First, we consider the weak coupling limit. 
In this case, the interface of the junction behaves as a hard wall. 
Because fields $\phi_{c}$ and $\phi_{f}$ are decoupled,~\cref{eqn:expval_1} is also decoupled as 
\begin{align} 
  \expval{e^{i\frac{\sqrt{\nu}}{2}p\phi_{c}^\mathrm{out}}e^{-i\frac{\sqrt{\nu}}{2}n\phi_{c}^\mathrm{in}}}_{c} 
  \expval{e^{iq\phi_{f}^\mathrm{out}}e^{-im\phi_{f}^\mathrm{in}}}_{f} 
  \expval{\eta_{p}^\mathrm{out}\eta_{n}^\mathrm{in}}_\mathrm{Ising} .
\end{align} 
For this to be nonzero, the condition 
\begin{align} 
  p=n \qc q=m 
\end{align} 
and 
\begin{align} 
  \eta_{p}^\mathrm{out}=\eta_{n}^\mathrm{in} 
\end{align} 
must hold. 
These results correspond to perfect reflections. 

Perturbative calculations in the weak coupling regime were reported in Ref.~\citenum{fradkin_2013} and the references therein. 
Similar to Refs.~\citenum{sandler_1998} and \citenum{fradkin_2013}, it is useful to introduce another basis: 
\begin{align} 
  \mqty(\phi_{1}\\\phi_{2})=\mqty(\cos\theta & \sin\theta \\ -\sin\theta & \cos\theta)\mqty(\phi_{c}\\\phi_{f})\qc 
  \tan\theta = \frac{1+\sqrt{\nu}}{1-\sqrt{\nu}}. 
\label{eqn:basis_transform} 
\end{align} 
Based on this new basis, the tunnel Hamiltonian can be rewritten as 
\begin{align} 
  H_\mathrm{T} = \int\dd{x}\Gamma\delta(x)\chi\cos\frac{1}{\sqrt{\nu^{\prime}}}\qty(\phi_{1}-\phi_{2}), 
\end{align} 
where 
\begin{align} 
  \nu^{\prime}=\frac{2\nu}{1+\nu}. 
\end{align} 
The scaling law for $\Gamma$ can be expressed as\cite{fradkin_2013} 
\begin{align} 
  \frac{\Gamma(\Lambda)}{\Gamma(\Lambda_0)}=\qty(\frac{\Lambda}{\Lambda_0})^{1/\nu^{\prime}-1/2}, 
\end{align} 
where $\Lambda_0$ and $\Lambda$ are the bare and renormalized cut-off values, respectively. 
At a high bias $eV>k_\mathrm{B}T$, the renormalized cut-off is proportional to $V$. 
The nonlinear $I$--$V$ characteristics are 
\begin{align} 
  I &\propto \Gamma(\Lambda)^2V \propto V^{1/\nu+1}
\end{align} 
and the $V$-dependence of the differential conductance is
\begin{align}
  \dv{I}{V}\propto V^{1/\nu}.
  \label{eqn:didv-weak}
\end{align}

\subsection*{Strong Coupling Limit} 
Next, we consider the strong coupling limit. 
This case corresponds to the weak coupling limit in the dual description;\cite{kane_1992a,kane_1992,furusaki_1993,nomura_2001,dec.chamon_1997,sandler_1998,fradkin_2013} in other words, this is the case of the model with a tunneling Hamiltonian with coupling $\tilde{\Gamma} = 0$:
\begin{align} 
  \tilde{H}_\mathrm{T}= 
  \int\dd{x}\tilde{\Gamma}\delta(x)\sigma\cos\frac{\sqrt{\nu^{\prime}}}{2}\qty(\tilde{\phi}_{1}-\tilde{\phi}_{2}). 
\end{align} 
Here, $\tilde{\phi}_{1}$ and $\tilde{\phi}_{2}$ are the dual fields $\phi_{1}$ and $\phi_{2}$, respectively.

The probability amplitude~\cref{eqn:expval_1} is decoupled in terms of the dual field, giving 
\begin{align} 
  \expval{e^{i\frac{\sqrt{\nu}}{2}p\tilde{\phi}_{c}^\mathrm{out}}e^{-i\frac{\sqrt{\nu}}{2}n\tilde{\phi}_{c}^\mathrm{in}}} 
  \expval{e^{iq\tilde{\phi}_{f}^\mathrm{out}}e^{-im\tilde{\phi}_{f}^\mathrm{in}}} 
  \expval{\eta_{p}^\mathrm{out}\eta_{n}^\mathrm{in}}_\mathrm{Ising} . 
  \label{eqn:expval_3} 
\end{align} 
Similar to the edge theory of the Laughlin state,\cite{sandler_1998} the dual and original fields are related to the incoming quasiparticles as
\begin{align} 
  \tilde{\phi}_{1}^\mathrm{in} = \phi_{1}^\mathrm{in}\qc 
  \tilde{\phi}_{2}^\mathrm{in} = \phi_{2}^\mathrm{in} 
\end{align} 
and for outgoing quasiparticles as 
\begin{align} 
  \tilde{\phi}_{1}^\mathrm{out} = \phi_{2}^\mathrm{out}\qc 
    \tilde{\phi}_{2}^\mathrm{out} = \phi_{1}^\mathrm{out}.
\end{align} 
Using these relations and~\cref{eqn:basis_transform}, the amplitude~\cref{eqn:expval_3} can be rewritten as 
\begin{align} 
  \begin{split} 
    &\expval{e^{i\qty(-p\frac{\sqrt{\nu}}{2}\sin2\theta+q\cos2\theta)\phi_{c}^\mathrm{out}}e^{-i\frac{\sqrt{\nu}}{2}n\phi_{c}^\mathrm{in}}}_c\\ 
    &\expval{e^{i\qty(p\frac{\sqrt{\nu}}{2}\cos2\theta+q\sin2\theta)\phi_{f}^\mathrm{out}}e^{-im\phi_{f}^\mathrm{in}}}_f 
    \expval{\eta_{p}^\mathrm{out}\eta_{n}^\mathrm{in}}_\mathrm{Ising} . 
  \end{split} 
\end{align} 
Then, we find that in the cases 
\begin{align} 
  \begin{split} 
    -p\frac{\sqrt{\nu}}{2}\sin2\theta+q\cos2\theta &= \frac{\sqrt{\nu}}{2}n \\ 
    p\frac{\sqrt{\nu}}{2}\cos2\theta+q\sin2\theta &= m 
  \end{split} 
\end{align} 
and 
\begin{align} 
  \eta_{p}^\mathrm{out} = \eta_{n}^\mathrm{in}, 
\end{align} 
the amplitude is nonzero. 
Our aim is to calculate the selection matrix $M$:\cite{sandler_1998}
\begin{align} 
  \begin{split} 
    \mqty(q\\p) &= M\mqty(m \\ n) \\ 
    &= \mqty(\dfrac{1-\nu}{1+\nu} & \dfrac{\nu}{1+\nu} \\ \dfrac{4}{1+\nu} & -\dfrac{1-\nu}{1+\nu})\mqty(m \\ n) .
  \end{split} 
\end{align} 
To ensure the relationship between the transmission and reflection probabilities of the electrons, let a large number $m$ of the electrons and zero quasiparticles be incident states in the presence of an external voltage. 
The reflection and transmission probabilities for electrons are given as 
\begin{align} 
  R_{e} &= \frac{q}{m} = M_{11} = \frac{1-\nu}{1+\nu}\\ 
  T_{e} &= \frac{(\nu e/2)p}{em} = \frac{\nu}{2}M_{21} = \frac{2\nu}{1+\nu}
\end{align} 
and similarly for quasiparticles as
\begin{align} 
  R_{\nu e/2} &= \frac{p}{n} = M_{22} = -\frac{1-\nu}{1+\nu}\\ 
  T_{\nu e/2} &= \frac{eq}{(\nu e/2)n} = \frac{2}{\nu}M_{12} = \frac{2}{1+\nu}.
\end{align} 
It should be noted that both relationships $T_{e} = 1 - R_{e}$ and 
$T_{\nu e/2} = 1 - R_{\nu e/2}$ hold. 
Thus, the two-terminal conductance can be calculated as 
\begin{align} 
  G &= \frac{e^2}{h}T_{e}=\nu\frac{e^2}{h}T_{\nu e/2} = \frac{e^2}{h}\frac{2\nu}{1+\nu}.
  \label{eqn:Landauer} 
\end{align} 

\begin{figure}[htb] 
\centering 
\includegraphics[width=0.7\linewidth]{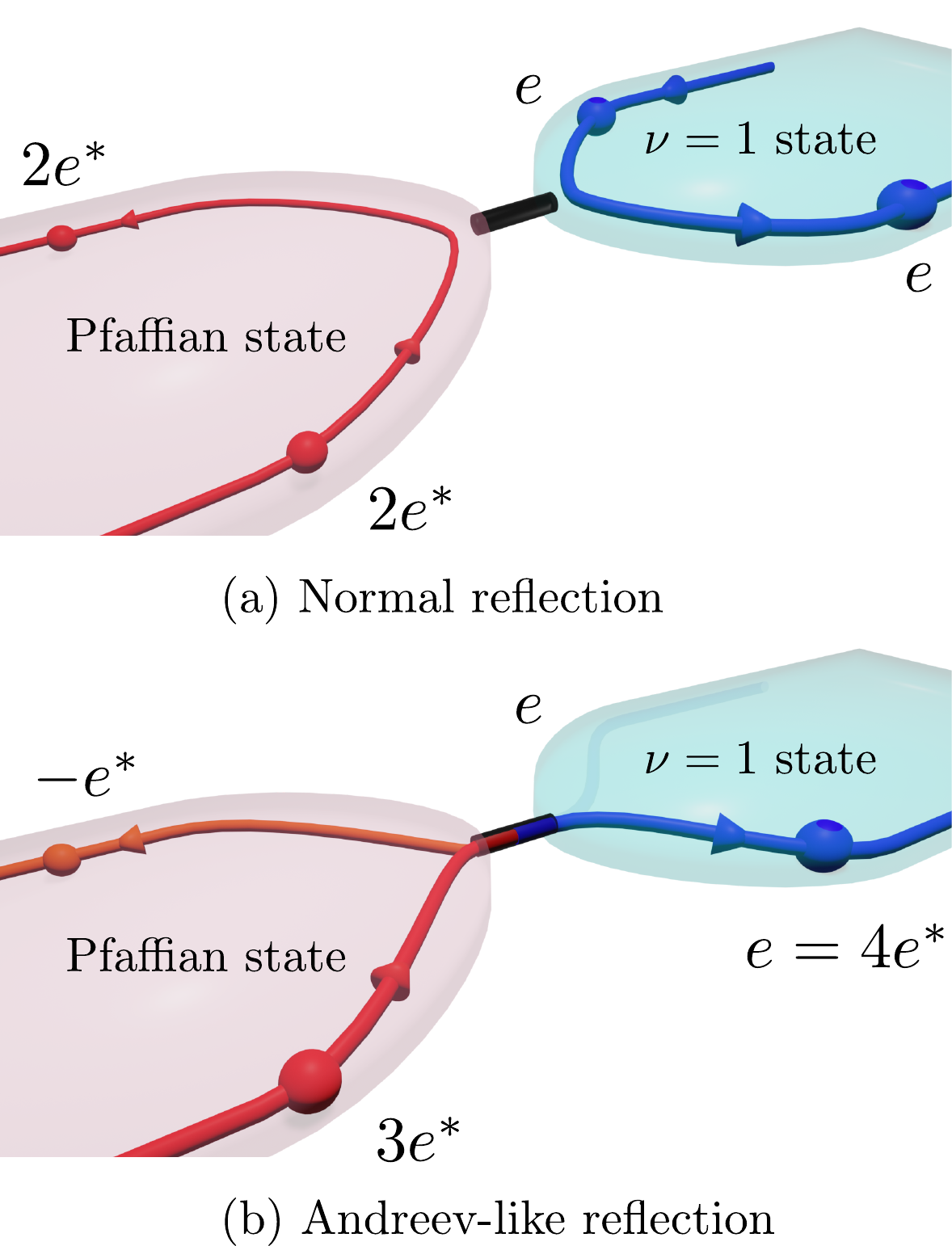} 
\caption{(Color online)
Two elementary processes at $\nu = 1/2$: (a) normal reflection and (b) Andreev-like reflection. The FQHL behaves as normal-metal. The elementary charge is $e^\ast = e/4$. 
} 
\label{fg: reflections} 
\end{figure}

At $\nu = \frac{1}{2}$ we find that there are two elementary processes: (a) a normal reflection of two $2e^{\ast}$ quasiparticles and an electron and (b) an Andreev-like reflection of $3$ quasiparticles with a total charge $3e^{\ast}$ in the initial and final states with one transmitted electron and one quasihole, as shown in~\cref{fg: reflections}. 
The latter process is analogous to an Andreev reflection at a normal-metal--superconductor (N--S) interface.\cite{andreev_1965,blonder_1982} 
Our selection rules derived from the dual theory are different from those considered in Ref.~\citenum{nomura_2001} and~\citenum{fendley_2007a}. 
For the Andreev reflection in the N--S junction, the energy gap of the S side is essential. 
In our case, we can consider the energy gap for the quasiparticle excitation as infinite on the IQHL side. 

\begin{figure}[htb] 
\includegraphics[width=\linewidth]{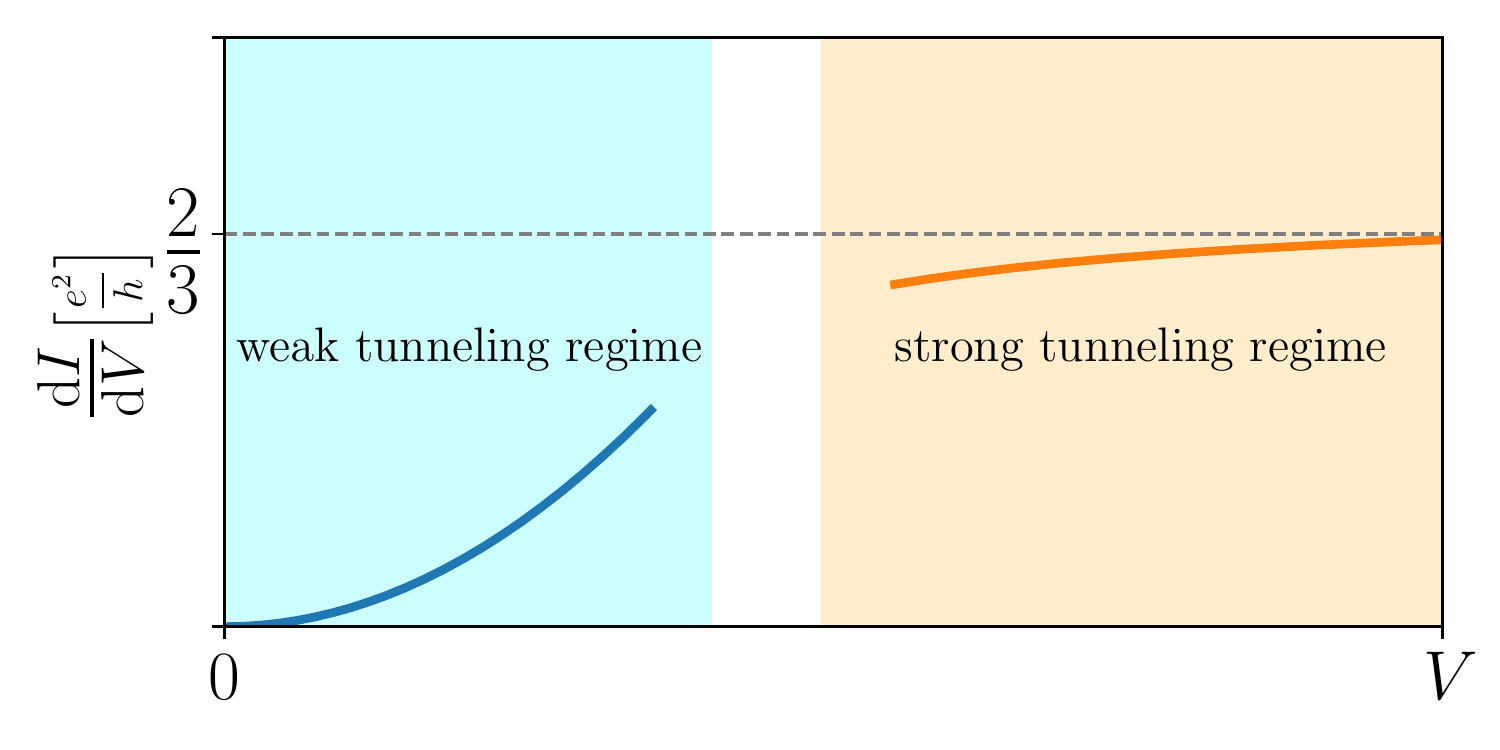} 
\caption{(Color online)
  Schematic behavior of the normalized differential conductance as a function of the bias voltage $V$. 
} 
\label{fg: graph_I_V} 
\end{figure}

Next, we consider the correction for $\tilde{\Gamma}$ in the strong coupling regime. 
By counting the conformal dimension, we obtain the renormalization group equation for $\tilde{\Gamma}$ as 
\begin{align} 
  \dv{\tilde{\Gamma}}{l}=\qty[1-\qty(\frac{\nu^{\prime}}{4}+\frac{1}{16})]\tilde{\Gamma}. 
\end{align} 
Then, the correction of the current is proportional to 
\begin{align} 
  \tilde{\Gamma}(\Lambda_0)^2\Lambda^{2\qty[\qty(\frac{\nu^{\prime}}{4}+\frac{1}{16})-1]}V. 
\end{align} 
At a high bias $eV > k_\mathrm{B}T$, the renormalized cut-off $\Lambda$ can be $V$. 
It can be assumed that the Landauer formula~\eqref{eqn:Landauer} holds in such a large-voltage regime. 
Therefore, we can consider a $\frac{1}{V}$-expansion in the strong coupling regime, where $\tilde{\Gamma}$ is small.
The $V$-dependence of the differential conductance is given by 
\begin{align} 
  \dv{I}{V}=\frac{e^2}{h}\qty(\frac{2\nu}{1+\nu})-\lambda V^{-\alpha}, 
  \label{eqn:didv_strong}
\end{align} 
where 
\begin{align} 
  \alpha = \frac{15}{8}-\frac{\nu}{1+\nu}, 
  \label{eqn:corr-term}
\end{align} 
and $\lambda$ is a constant proportional to $\tilde{\Gamma}(\Lambda_0)^2$. 
The behavior of the differential conductance in both the weak and strong tunneling regimes is illustrated in~\cref{fg: graph_I_V}. 

\begin{figure}[htb] 
\centering 
\includegraphics[width=0.8\linewidth]{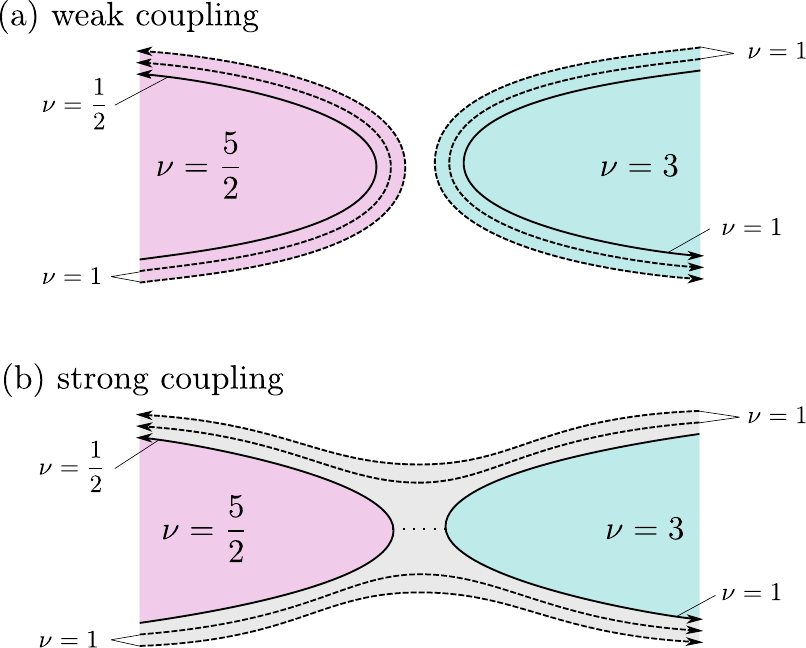} 
\caption{(Color online)
  Schematic illustration of the coupling in a feasible model. 
} 
\label{fg: exist_model} 
\end{figure}

Finally, to compare the above model with experimental conditions, we extend it for the system with FQHL with $\nu=5/2$ and IQHL with $\nu=3$, as shown in~\cref{fg: exist_model}. 
The FQHL with $\nu=5/2$ can be regarded as a system in which $\nu=2=1+1$ and $\nu=1/2$ coexist.\cite{fradkin_2013} 
The usual coupling occurs between each $\nu=2$ liquid, and in this study the extra $\nu=1/2$ and $\nu=1$ liquids occur. 
Thus, the differential conductance value increases owing to the $\nu=2$ coupling term. 

In this study, we considered the Pfaffian FQHL--IQHL junction. 
In the strong coupling limit, the selection rule shows a normal reflection for quasiparticles with charge $2e^{\ast} = \nu e$ and an Andreev-like reflection for elementary quasiparticles with charge $e^{\ast} = \nu e/2$. 
The current--voltage characteristics were calculated in the weak and strong coupling regimes. 
In the weak tunneling limit, $\dd{I}/\dd{V}$ is proportional to $V^{1/\nu}$. 
By contrast, in the strong tunneling limit, $\dd{I}/\dd{V}$ is expressed by $(e^{2}/h)2\nu/(1+\nu)$ with the correction term $-\lambda V^{-\alpha}$.

\begin{acknowledgment} 
R.O. was supported by JSPS KAKENHI (Grant No. JP21J15526). 
R.N. was supported by JSPS KAKENHI (Grant No. JP17K17604).
T.Y. was supported by JSPS KAKENHI (Grant N0. JP30578216) and the JSPS-EPSRC Core-to-Core program "Oxide Superspin". 
Y.T. was supported by Scientific Research (A) (KAKENHI Grant No. JP20H00131) and Scientific Research (B) (KAKENHI Grant No. JP20H01857). 
K.N. was supported by KAKENHI (Grant No. JP20H01830) and CREST, Japan Science and Technology Agency (Grant No. JPMJCR18T2). 

\end{acknowledgment} 


\bibliographystyle{jpsj} 
\bibliography{17911.bbl} 

\end{document}